\documentstyle[12pt,psfig]{article}
\begin{document}
\setcounter{page}{1}

  \hspace{16.6em}
 {\Large   SUNY -  NTG    96-40
  }
\vspace*{0.8cm}

  \begin{center}
{\bf  \large
Elliptical flow -- a signature for early pressure 
     \\
in ultrarelativistic nucleus-nucleus collisions
       }
\vspace*{0.4cm}\ \\
           H.\ Sorge
 \footnote{
  E-mail: sorge@nuclear.physics.sunysb.edu
          }
\vspace*{0.4cm}\ \\
 Physics  Department,
\vspace*{0.4cm}\ \\
  State University of New York at Stony Brook, NY 11794-3800

  \end{center}

\begin{abstract}
Elliptical energy flow patterns in non-central Au(11.7AGeV) on Au reactions
have been studied employing the RQMD model. The strength of these azimuthal 
asymmetries is calculated  comparing the results in two different modes of 
RQMD (mean field and cascade). It is found that the elliptical flow which is 
readily observable with current experimental detectors may help to distinguish 
different reasonable expansion scenarios for baryon-dense matter. The final 
asymmetries are very sensitive to the pressure at maximum compression, because 
they involve a partial cancelation between early squeeze-out and subsequent
flow in the reaction plane. This cancelation can be expected to occur in a 
broad energy region covered  by the current heavy ion fixed-target programs 
at BNL and at CERN.
\end{abstract}
 
A  primary goal of current  heavy-ion physics utilizing  beams at
 ultrarelativistic energies is the creation and  observation of the 
quark-gluon plasma (QGP), a phase in which quarks and gluons
are deconfined and chiral symmetry has been restored.
The extraction of flow signatures from experimental data
has found considerable
interest recently amid  present ambiguities concerning the QGP formation.
Since collective flows are driven by pressure gradients,
their measurement provides a   diagnostic tool to study the 
transient pressure in these reactions.  
A first-order phase transition is generically associated
with  the presence of a `softest point' in the equation of state.
The tendency of matter to expand on account of its internal pressure is
reduced  in the transition region \cite{shuryak79,transhydro}.
 It has been predicted that a softening of the equation
 of state can be deduced from experimental measurements, 
 e.g.\ by observing a minimum in the directed nucleon flow 
\cite{bravina,rischkea} or maximum in the life time of the created
 hot medium \cite{hung95} as a function of beam energy.
It should be also kept in mind that there is no guidance
from lattice QCD about the physics of the  quark-hadron
transition at nonzero baryon density which is relevant for
today's heavy ion experiments  with beam energies up to 200 AGeV \cite{qmatter}.
Experimental information on the pressure generated in the collision region is
therefore clearly warranted. 
     
The so-called directed transverse flow in the reaction plane
and the  squeeze-out, a larger flow of nucleons in the
central region out-of-plane, have been studied extensively at energies
around 1 AGeV \cite{partlan95}, available at SIS-GSI and before
at the BEVALAC.
Recently, direct signals of  transverse flow have been reported 
both at AGS and the higher CERN energies.
 E877 has analysed the azimuthal distributions of transverse energy and
charged particle multiplicity using the technique of
 Fourier decomposition \cite{flow814}.
 Evidence for a nonvanishing first and second moment
 of the azimuthal distributions has been found \cite{qm96-e877}.
The first moment 
reflects the directed transverse flow and can be used to determine 
the reaction plane. A nonvanishing second moment signals the
elliptical deformation of the flow tensor.
The preliminary E877 indicate  that the main
flow direction is parallel to the impact parameter,
not orthogonal as expected for a squeeze-out.
This was predicted by   Ollitraut some time ago for
collisions at sufficiently high energies  \cite{ollitrault}.
NA49 has reported preliminary data on  elliptical transverse energy
flow patterns for semi-peripheral Pb(158AGeV) on Pb reactions,
too \cite{qm96-na49}. Although NA49 is not able to determine a reaction plane
a clear signal above statistical fluctuations
has been found by correlating the prefered directions of emission
in neighboring pseudo-rapidity windows.
   
In this letter, I focus on  elliptical flow in the central
(pseudo-)rapidity region in order  to extract information
 on the pressure in the collision zone.
  The highest energy and baryon densities
 are achieved in the central region for present experiments.
 Particles in the central region are presumably less affected by
 nonequilibrium physics. In contrast,
the directed flow  results mostly from  a 
transverse momentum anticorrelation between baryons
near to original projectile and target rapidity.
Therefore it receives a strong  preequilibrium contribution.  
     
Measurement of elliptical flow  patterns provides a unique opportunity to 
study  the pressure which is generated very {\em early}  in the reaction. 
The basic idea presented here is very simple.
At low beam energy (around 1AGeV), matter escapes preferentially orthogonal to
the reaction plane which is spanned by the beam axis and the impact parameter.
The spectator nucleons block the path of participant
hadrons which  try to escape from the collision zone.
 This is the observed squeeze-out effect.
At ultrahigh energies, a larger flow of participants
{\em in} the reaction plane can be expected 
and is confirmed by RQMD calculations for RHIC energy \cite{rhicss96}.
Since the passage time of spectator nucleons from
projectile and target shrinks with the Lorentz factor
gamma, produced particles do not interact with spectators.
The almond-shaped geometry of the overlap region
in the reaction plane clearly favors preferential emission
parallel to the impact parameter.
An interesting situation emerges for collision energies 
between the low and the ultrahigh energies, basically
the whole energy region covered by present fixed target experiments
at BNL-AGS and CERN-SPS (2-200AGeV).
Taking collisions of equal mass nuclei moving with the speed of light, 
the passage time of projectile and target spectators is approximately
given by $ 2 R_A/ (\gamma c)$ (numerically 5.4 fm/c at 12AGeV and 1.4 
fm/c at 160AGeV for the heaviest systems).
Such time scale neither covers the whole reaction time
nor becomes irrelevant  at these intermediate energies.
As a consequence, the centrally produced matter is initially 
 squeezed-out orthogonal to the reaction plane. After
the spectator material has disappeared, the `confining' spectator walls
suddenly have vanished. Lateron,  the  geometry of the central region
favors central flow  parallel to the impact parameter vector.
The orientation of the final azimuthal asymmetry 
in particle, momentum and energy flow is chiefly determined by
the relative strength of the pressures
during the initial passage as compared to the later expansion time.
The full power of this analysis becomes apparent  if 
the measurement of elliptical flow patterns is combined  with
measurements of the average flow in transverse
directions, the so-called radial flow. While the elliptical flow is 
influenced by the difference between early and late pressures the
average transverse flow reflects the time integral  of these pressures.
It is the main virtue of such kind of analysis  to
shed some light on the early pressure which is practically
unknown for baryon-dense matter. The later expansion stage
characterized by baryon densities around 0-2 times ground state
density is clearly more constrained from known nuclear physics
and ongoing heavy ion studies at BEVALAC-SIS energies.

In the following, I am going to employ a transport model,
 relativistic quantum molecular dynamics (RQMD) \cite{SOR95}, to calculate the
 azimuthal asymmetries in the energy deposition 
for Au(11.7AGeV) on Au collisions in non-central collisions.
I shall discuss whether  the predicted  partial cancelation of 
 in-plane and out-of-plane emission 
can be  utilized to quantitatively distinguish different
dynamical evolution scenarios
for highly compressed baryon-rich matter. 
RQMD is constructed  as a  Monte-Carlo code which generates
complete events under prescribed conditions
(masses of the colliding nuclei, impact parameter, beam energy).
RQMD is based on string and resonance excitations in
the primary collisions of nucleons from target and projectile.
Overlapping color strings may fuse into so-called ropes.
Subsequently, the fragmentation products from rope, string
and resonance decays  interact with each other and the
original nucleons, mostly  via binary collisions. 
These interactions drive the system towards equilibration
\cite{sorplb95} and are responsible that collective flow
develops, even in the preequilibrium stage.
RQMD contains some option which allows to vary the
pressure in the high-density state and to study its influence
on final-state observables.
 Baryons acquire effective masses   and thus may  eperience forces 
   if they are surrounded by  other baryons  \cite{SOR89}.
  The effective masses are generated by introducing
   Lorentz-invariant quasi-potentials into the mass-shell constraints
   for the momenta which simulate the effect of `mean fields'.
  There are no potential-type interactions in the
 so-called   cascade mode  of RQMD. 
  In this mode,  the equilibrium pressure is  simply an ideal gas  of
   hadrons and resonances which are explicitly  propagated and therefore
  contribute to the pressure.
 The resulting  equation of state in the cascade mode of RQMD
  is very similar to the one  calculated and plotted by the Bern group in
 Ref.~\cite{bebie92}, because
 the spectrum of included resonance states is nearly the same. 
\footnote{
 It should  be noted that 
 propagating strings  modify  the equation of state as well.
 This  correction  is small, however, in equilibrium at
 relevant temperatures around 150 MeV.
 Although it is not realized in RQMD  the collision term in the equations of
 motion may  contribute to the equilibrium pressure, in   principle.
 For instance, repulsive trajectories are 
 selected for colliding baryons  with some probability
 in new versions of the ARC model \cite{arc96}. In Refs.~\cite{rqmd92,rqmdmf}
 it was concluded that a cascade lacks some pressure
  in comparison to experimental data for $AA$ collisions
 at 10-15 AGeV. This result was seemingly contradicted
  by the findings in \cite{arc96},  based on ARC calculations.  However, 
  the latter calculations contain some nonideal gas
  pressure contributions from the repulsive trajectory prescription.
}
   
RQMD results for ultrarelativistic nucleus-nucleus collisions 
have been  compared to measurements by most
major experimental collaborations \cite{qmatter}, showing usually 
reasonable or good agreement. Various experimental data 
-- e.g.\ directed and total transverse momenta
  \cite{qm96-e877,qm95-e866} --  which have been taken
in the AGS energy region around 12 AGeV seem to hint that 
the generated pressure in RQMD is too `soft' if it is used in its cascade mode.
Therefore we compare here the result obtained in the cascade mode
with a calculation in which the quasi-potentials generate
additional pressure due to repulsion at baryon densities larger
than ground state density. Potential parameters  have been 
selected for RQMD (version 2.3)
 which bring the generated transverse momenta  in
agreement with available  data similarly as it was done in \cite{rqmd92}.
 Since  a first order phase transition from a resonance gas
into a QGP would   `soften'
 the equation of state, its inclusion in RQMD would act
in the  opposite direction to
 repulsive mean fields. It is  worthwile  to study a   softening 
  as a possible consequence of the hadron-quark
 transition in more detail in the RQMD framework. 
  
\begin{figure}[h]
\vspace{-3.0cm}

\centerline{\hbox{
\psfig{figure=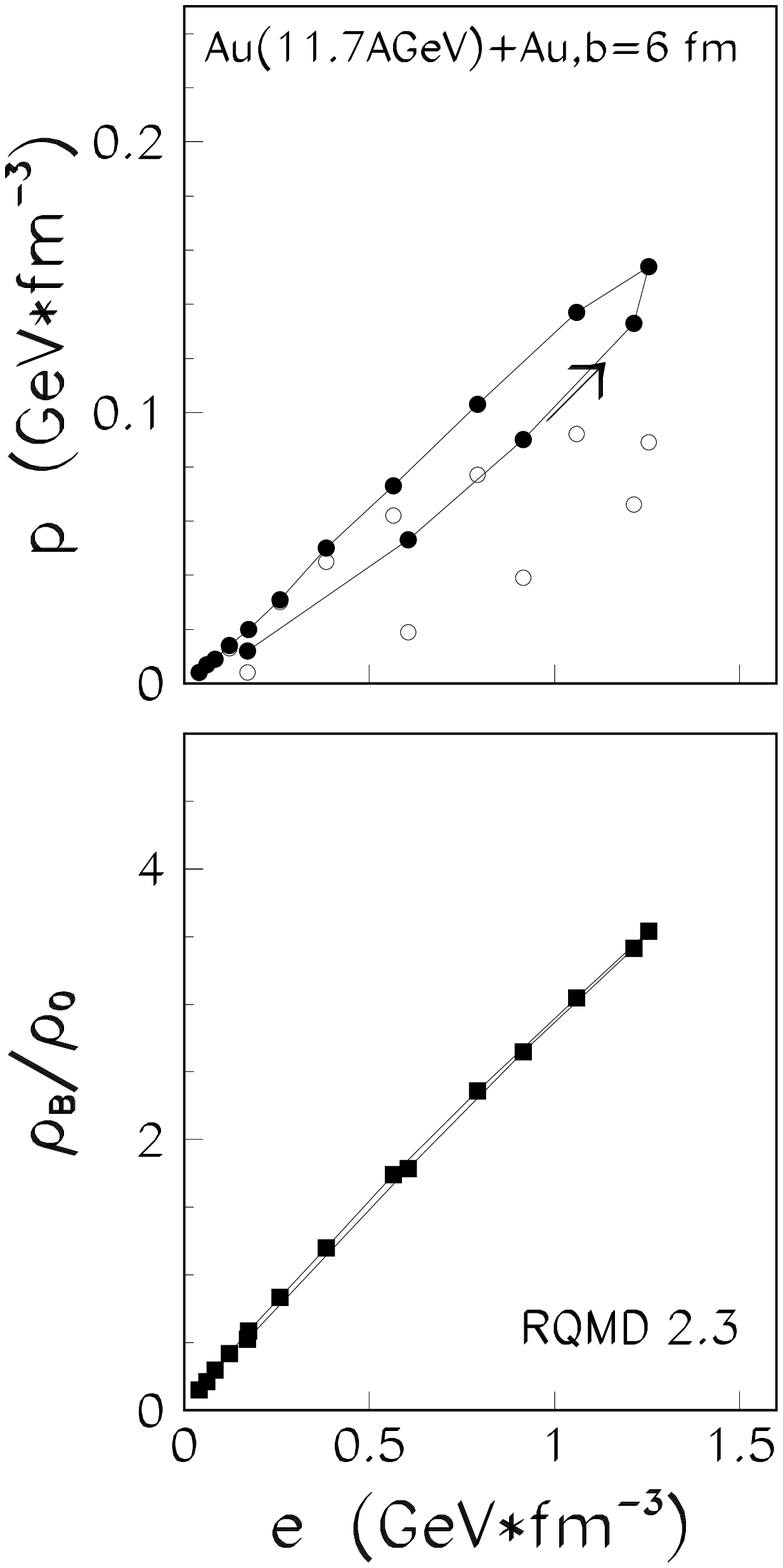,width=10cm,height=20cm}}}

\caption
[
 ]
{
 \label{figure1}
}
\end{figure}

 In the following, the question will be addressed how sensitive
 the azimuthal asymmetry of energy flow in the central
  rapidity region is to the pressure in the early and in the  late stage.
  The terms `early' and `late' are defined with respect to the
   passage time of target and projectile spectators.
 For this purpose, I have analysed the evolution of the
  pressure in a particular reaction Au(11.7AGeV) on Au at an  impact parameter
 of 6 fm.  The equations of motion  contain a contribution from
 quasi-potentials.   This has to be taken into account in the definition of
 the non-equilibrium pressure.   
 The virial theorem has been applied to define
  the  pressure in each of the three space directions
 via the equation
\begin{equation}
  \label{virial}
   P^i \cdot V = \left\langle 
                \sum _{M} 
                  \left(
                  {\bf p}^i(M) {\bf v}^i(M) +
                 \sum _{N}  {\bf F}^i_{MN} {\bf r}^i(M) 
                \right)  
                \right\rangle
                     \qquad  i=1,2,3
                     \quad .
\end{equation}
   ${\bf p}$,  ${\bf v}$ denote the hadron's momentum and
    velocity, ${\bf F}_{MN}$ the force which baryon 
    $N$ exercizes on  $M$.  The summations over $N$, respectively  $M$,
   include only hadrons inside a cylindrical
  volume $V=2\pi (x R_{A})^3/\gamma$   centered at the origin (with $x$=0.3).
  Ingoing nucleons which have not collided yet are not included
   in the evaluation of the pressure.
 The rhs of eq.~(\ref{virial}) has been evaluated for
  approximately 400 events in the mean field mode of RQMD.
   An event average has been taken
  (indicated by the $\langle \rangle$ symbols).
 Analogously, local energy and baryon densities  ($e$ and $\rho_B$) have been
  calculated.  One should keep in mind that
  the system is initially in non-equlibrium.
  This means that the pressure along the beam axis is the largest at this stage.
   Since no nonstatistical difference between the two transverse pressure
 components     was found, their average was taken and is dubbed
  `pressure' in the following.
 The CMS time evolution of  pressure, local energy
 and baryon density  can be seen from   Fig.~\ref{figure1}.
 The upper part of Fig.~\ref{figure1} shows the evolution
   in the $e$-$p$ plane, the lower part in the  $e$-$\rho_B$  plane.
   The symbols represent the values of these quantities, with 
    time increasing in steps of 1 fm/c each, starting  1 fm/c after
   the two nuclei have touched each other.
 It becomes apparent from this figure that the passage time
  is practically identical with the time of maximum compression,
  either of energy or baryon number.
 The energy and baryon density at maximum  compression
  are approximately 1.3 GeV/fm$^3$ and 3.5 $\rho_0$, respectively. 
  It is noteworthy that these values are close to the region
  for which a phase transition  into a QGP is usually
  expected. 
  The pressure in the expansion stage is somewhat larger
   than in the compression stage. This reflects the
   presence of a strong preequilibrium component 
   which tends to soften the transverse pressure.
   (Remember that longitudinal momentum acts like a mass
   term with respect to      velocity in transverse directions.)
In Fig.~\ref{figure1} the contribution from the kinetic part of the     
pressure (the first term on the rhs of eq.~(\ref{virial}))
 is also shown (open symbols). Roughly, the kinetic
 and the potential part contribute equally at the time of 
 maximum compression.
  
\begin{figure}[h]
\vspace{-3.0cm}

\centerline{\hbox{
\psfig{figure=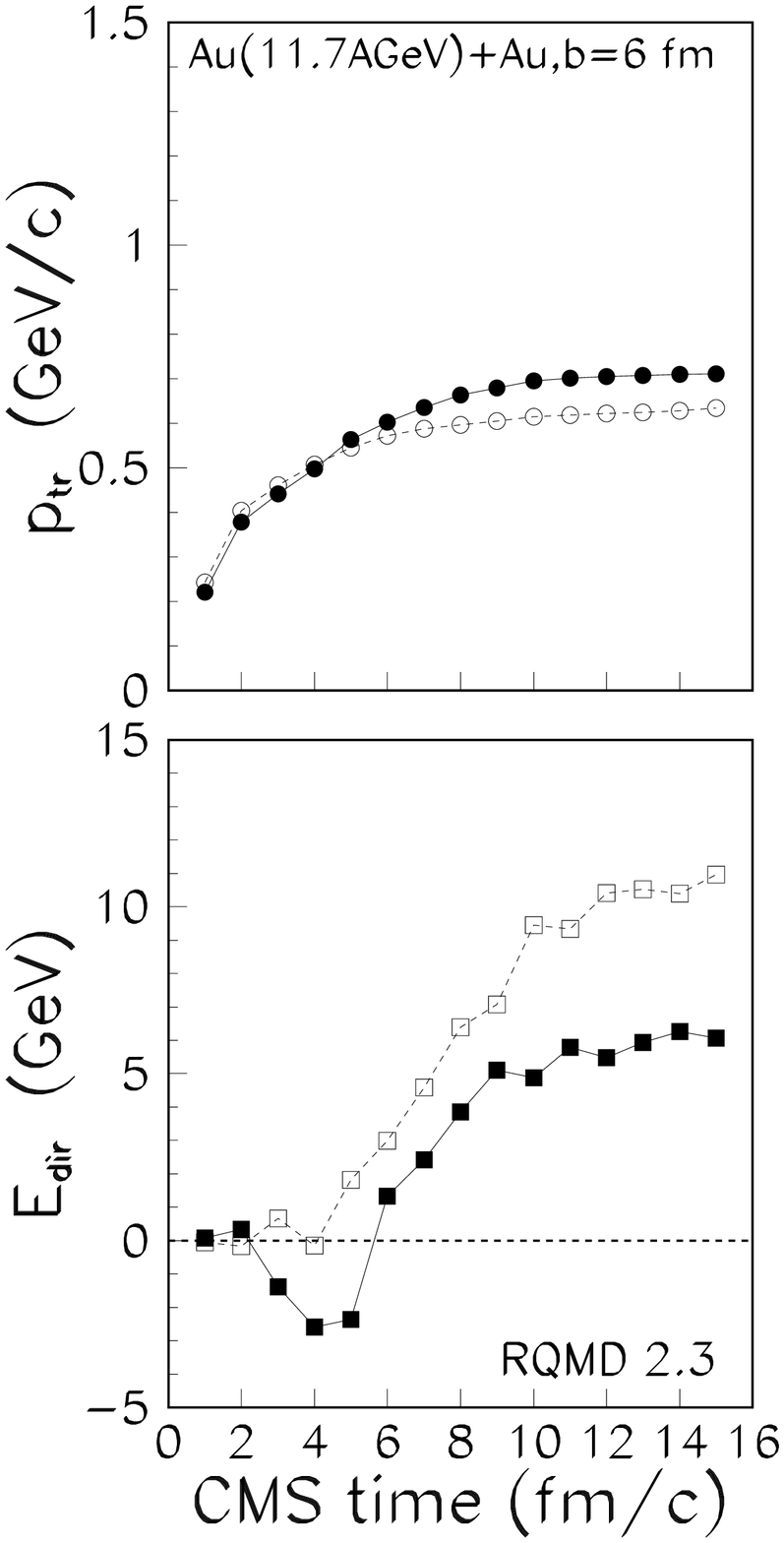,width=10cm,height=20cm}}}

\caption
[
 ]
{
 \label{figure2}
}
\end{figure}

\begin{figure}[h]
\vspace{-3.0cm}

\centerline{\hbox{
\psfig{figure=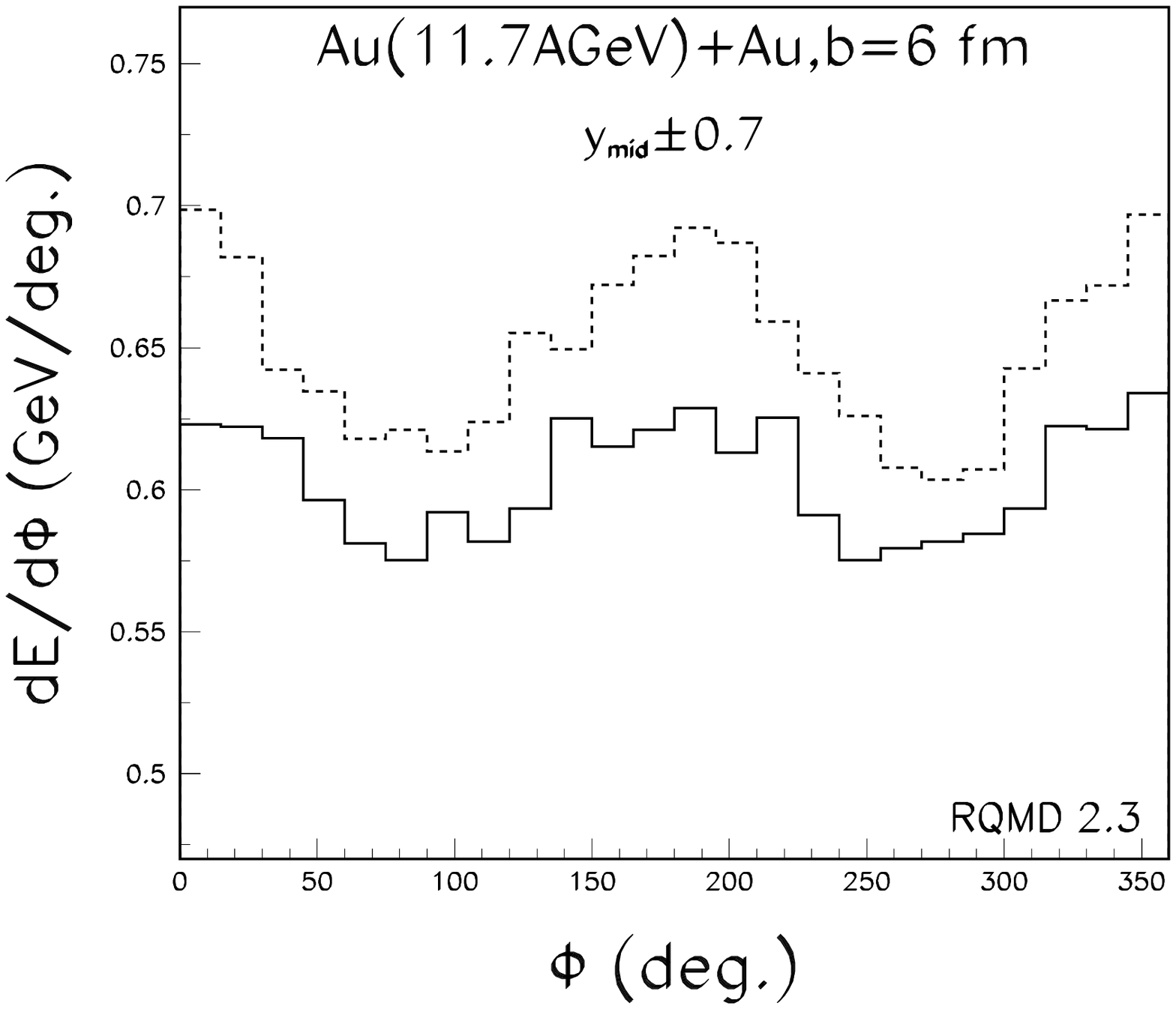,width=20cm,height=20cm}}}

\caption
[
 ]
{
 \label{figure3}
}
\end{figure}

The azimuthal asymmetry in the energy flow can be quantified
 by defining the following variable:
\begin{equation}
  \label{endir}
   E_{dir} = 
                \sum _{M}  E(M) \cdot 
                   \mbox{sgn}(\phi)
                     \quad .
\end{equation}
    The summation over $M$  includes  hadrons only
   within some rapidity cut around center-of-mass rapidity,
    set here arbitrarily to $\pm$0.7.
   $\phi $ is defined as  the angle of a hadron's momentum
   with respect to the impact parameter vector.
   $\mbox{sgn}(\phi)$ is defined to be +1 in  the 
  cones with opening angle of 45$^o$ around $\phi$= 0
   and 180$^o$, -1 elsewhere.
Fig.~\ref{figure2} displays the time evolution of   $E_{dir}$ in the CMS.
   The time dependence of   $E_{dir}$ shows the behaviour    
  as expected from the qualitative  discussion above. 
   In the time span right before maximum compression,
   $E_{dir}$ acquires negative values  (squeeze-out),
   because the pressure mostly from the repulsive potentials 
    pushes the hadrons against the confining spectator material
  in the reaction plane and into the vacuum orthogonal to it.
   After the spectators are gone (t$>$5 fm) 
     hadrons are pushed preferentially (anti-)parallel to the
    impact parameter vector. In course, 
   $E_{dir}$ gets positive contributions with increasing time.
   Finally, the  `in-plane flow' effect dominates 
   for this reaction. Therefore the major energy flow axis
    is parallel to $\vec{b}$. For comparison, 
Fig.~\ref{figure2} shows the evolution of the same quantity, but
   calculated in the RQMD cascade mode.
  Due to the preequilibrium effect, the effective transverse equation of
 state is ultrasoft. There is no visible squeeze-out 
  present at the early times. The pressure at later times
    is smaller than  in the  mean field mode. However, 
  the final azimuthal asymmetry expressed by its $E_{dir}$
   value is approximately 60 percent larger, because
  the initial squeeze-out is absent.
  This is also visible from  Fig.~\ref{figure3}
   in which the full $\phi $ dependence of the energy  flow
   is shown, calculated with RQMD in the two different modes.
  I conclude that the azimuthal asymmetries are a tool of utmost importance 
  to gain information on  the pressure in the high density stage. 
   A factor 2 effect in the early pressure is not washed out 
   in the later evolution but shows up with practically
   undiluted strength in the different azimuthal asymmetries
    which are generated with and without mean field-type
   interactions.

 In addition,  Fig.~\ref{figure2}  shows the evolution of the 
  average nucleon transverse momentum 
   (in the same rapidity window as for $E_{dir}$).
   Again, the results obtained in the mean field mode are compared to
   those of the cascade mode. The mean field effect is noticeably smaller
   for the inclusive single-particle observable $p_{tr}$ (10 percent).
   Similar observations have been made based on calculations
   with the transport code ART comparing
    the sensitivity of directed and total transverse
    momenta to mean fields \cite{art95}.
  The transverse momentum  measurements favor the inclusion
   of repulsive potentials in RQMD. 
    On the other side,  those are most relevant at
     time of maximum compression. Thus they necessarily weaken
    the observable azimuthal asymmetry 
   (cf.\ Figs.~\ref{figure2} and ~\ref{figure3}). 
   This RQMD  prediction is readily testable using the present
    experimental set-up of the E877 group. 
   We take  the {\em qualitative } agreement with preliminary E877 data 
     which show major flow parallel to the impact parameter
   as an encouring sign. The model seems to be  sensible
    enough to address the question of the 
   centrally produced pressure quantitatively.

Summarizing,   I have analysed  calculations
 which were done with the RQMD model for the reaction
 Au(11.7AGeV) on Au at   impact parameter 6 fm.
 Azimuthal asymmetries in the energy flow  have been studied 
 in relation to the transient pressures in the central  collision zone.
  Employing repulsive mean fields, the elliptical flow
  is   initially oriented orthogonal to the impact parameter
   and acquires  lateron parallel contributions.
   While the total pressure integral can be constrained if not
   determined by other means, the azimuthal asymmetry is
    sensitive to the difference of the early and late pressures.
  I believe that this effect is present in a wide
   range of beam energies covered by today's fixed target experiments.
   Although I have studied here only one particular reaction,
  the most important findings will hold true very generally.
   I suppose that the effect  can be qualitatively 
   reproduced by other -- hydrodynamical or transport  -- approaches.
   This means in turn that the    elliptical flow provides an extremely useful
   tool in a joint effort of  experimentalists 
    (e.g.\  E877  and E895 \cite{e895}) 
   and theorists 
    to  gain quantitative information on  the early 
   pressure in the most dense -- the central -- collision region.
   Such a tool has been lacking so far, because the directed flow
    is concentrated in the projectile and target fragmentation regions.
   The partial cancelation of the azimuthal asymmetry in the flow of
    energy (and momentum and presumably particles)   with  time opens 
 up  possibilities for a rich experimental and theoretical research program.
   It is obvious from the presented calculations that the
   transition point at which  the squeeze-out disappears
    and is replaced by an in-plane flow will   strongly
   depend on the strength of the early pressure.
 Non-central asymmetric collisions and variations of  centrality triggers
  could provide much more detailed and additional information.
  As it is customary in the studies of the directed flow one
   can impose cuts on momenta or on particle species
   (e.g., kaons versus protons) in order
  to (de-)emphasize the pressure contribution from the early high density
   stage. In particular, it should be possible to considerably narrow down the
    pressure in  matter with  baryon densities of several times 
    ground state density  for which only educated guesses exist so far.

The author thanks E.\ Shuryak  for useful discussions. 
This work has been
supported by DOE grant No. DE-FG02-88ER40388.

\newpage

{\noindent  \LARGE   Figure Captions:}
\vspace{0.6cm}

{\noindent \large Figure 1: }

{\noindent       
  CMS time evolution of transverse  pressure $p$, local energy
 and baryon density  ($e$, respectively $\rho_B/\rho_0$) 
   in the collision center
  for the reaction
 Au(11.7AGeV) on Au at  impact parameter  6 fm.
  The results were obtained using the RQMD model 
   (version 2.3) in mean field mode.
 The upper part shows the evolution
   in the $e$-$p$ plane, the lower part in the  $e$-$\rho_B$  plane.
   The symbols represent the values of these quantities, with 
    time increasing in steps of 1 fm/c each, starting  1 fm/c after
   the two nuclei have touched each other.
  The time direction is indicated by an arrow.
  The contribution from the kinetic part of the     
 pressure alone is also  displayed (open circles). 
}
\vspace{0.2cm}

{\noindent \large Figure 2: }

{\noindent       
  CMS time evolution of the  average transverse momentum of nucleons
   $p_{tr}$
   (upper part of figure) and  $E_{dir}$ (bottom).
  The variable $E_{dir}$ is defined in               
  eq.~(\ref{endir}) and   is related to the
   direction of energy flow with respect to the
   impact parameter vector.
   Positive $E_{dir}$ values correspond to 
   major in-plane flow and negative values
    to out-of-plane flow (squeeze-out).
   The results were obtained for the same system as in Fig.~1.
    Closed (open) symbols refer to 
     the calculation in RQMD mean-field (cascade) mode.
    A  rapidity  cut $\pm$0.7 around  center-of-mass rapidity
    was imposed on hadrons which are included here.
}
\vspace{0.2cm}

{\noindent \large Figure 3: }

{\noindent       
   Differential energy flow distribution $dE/d\phi$ as a function
   of the  angle  with respect to the
   impact parameter vector $\phi $. 
   The results were obtained for the same system as in Figs.~1
   and 2 and the same acceptance cuts as in Fig.~2.
    Straight (dashed) line histogram refers to 
     the calculation in RQMD mean-field (cascade) mode.
}

\newpage  

\newpage  

\newpage  

\end{document}